\documentclass[journal]{IEEEtran}
\usepackage{amsmath,amsfonts}
\usepackage{algorithmic}
\usepackage{algorithm}
\usepackage{array}
\usepackage[caption=false,font=normalsize,labelfont=sf,textfont=sf]{subfig}
\usepackage{textcomp}
\usepackage{stfloats}
\usepackage{url}
\usepackage{verbatim}
\usepackage{graphicx}
\usepackage{cite}

\usepackage{amsfonts}
\usepackage{multirow}
\usepackage{booktabs}
\usepackage{longtable}
\usepackage{amsmath}

\usepackage{subfig}
\usepackage{ulem}

\usepackage{tikz,xcolor,hyperref}
\hypersetup{
    colorlinks=false,
    hidelinks
    }
\definecolor{lime}{HTML}{A6CE39}
\DeclareRobustCommand{\orcidicon}{%
	\begin{tikzpicture}
	\draw[lime, fill=lime] (0,0) 
	circle [radius=0.16] 
	node[white] {{\fontfamily{qag}\selectfont \tiny ID}};
	\draw[white, fill=white] (-0.0625,0.095) 
	circle [radius=0.007];
	\end{tikzpicture}
	\hspace{-2mm}
}

\foreach \x in {A, ..., Z}{%
	\expandafter\xdef\csname orcid\x\endcsname{\noexpand\href{https://orcid.org/\csname orcidauthor\x\endcsname}{\noexpand\orcidicon}}
}

\usepackage{pifont}
\newcommand{\xmark}{\ding{53}}

\begin{document}

\title{Machine Perception-Driven Image Compression:\\ A Layered Generative Approach}

\author{Yuefeng Zhang, Chuanmin Jia, Jianhui Chang, Siwei Ma\orcidD{},~\IEEEmembership{Senior Member,~IEEE}
}



\maketitle

\begin{abstract}
In this age of information, images are a critical medium for storing and transmitting information.
With the rapid growth of image data amount, visual compression and visual data perception are two important research topics attracting a lot attention. However, those two topics are rarely discussed together and follow separate research path. Due to the compact compressed domain representation offered by learning-based image compression methods, there exists possibility to have one stream targeting both efficient data storage and compression, and machine perception tasks. 
In this paper, we propose a layered generative image compression model achieving high human vision-oriented image reconstructed quality, even at extreme compression ratios.
To obtain analysis efficiency and flexibility, a task-agnostic learning-based compression model is proposed, which effectively supports various compressed domain-based analytical tasks while reserves outstanding reconstructed perceptual quality, compared with traditional and learning-based codecs.
In addition, joint optimization schedule is adopted to acquire best balance point among compression ratio, reconstructed image quality, and downstream perception performance.
Experimental results verify that our proposed compressed domain-based multi-task analysis method can achieve comparable analysis results against the RGB image-based methods with up to $99.6$\% bit rate saving (i.e., compared with taking original RGB image as the analysis model input). 
The practical ability of our model is further justified from model size and information fidelity aspects.

 
\end{abstract}

\begin{IEEEkeywords}
Image compression, machine perception, generative adversarial network, visual applications.
\end{IEEEkeywords}

\section{Introduction}

\IEEEPARstart{V}{isual} data is created at an incredible speed due to the great progress in electronic image acquisition area. On the one hand, data compression is in need to efficiently store and transmit mass visual data. On the other hand, high-level semantic understanding is required to translate pixel-level signals into machine-oriented knowledge. In the past, those two domains are seldom studied simultaneously due to the different targets between two domains \cite{Yang2021VideoCF}.

The increasing volume of visual data emphasizes the need for not only storage and transmission, but also machine perception. Apart from the goal of effectively reconstructing pixels for human vision, image/video coding standards are beginning to pay attention to high-level machine perception and pattern recognition tasks. To support large scale image/video retrieval and analysis, MPEG introduced standards: compact descriptor for visual search (CDVS) (ISO/IEC15938-13) \cite{duan2016overview} in Sep.2015 and compact descriptors for video analysis (CDVA) (ISO/IEC15938-14) \cite{duan2019compact} in July 2019. At the same time, JPEG AI \cite{Joao2020laerningbased} is launched in March 2019 targeting both human vision and computer vision tasks.

Video coding for machine (VCM) \cite{duan2015overview} or intelligent coding \cite{alvar2019multi} is proposed aiming at building an efficient joint visual compression and analysis framework. Those methods can be categorised into two. The first one \cite{gueguen2018faster,ehrlich2019deep} intends to make the reconstructed image more suitable for machine-oriented tasks. Most methods of this category are built on traditional codecs through introducing analysis-related information. The second one \cite{torfason2018towards,liu2021semantics} conducts visual analysis directly on the intermediate compressed features. In this category, learning-based compression methods have the advantage that their deep feature is full of compact semantic information, which traditional codecs does not have.

Taking the advantage of deep learning, learning-based compression methods begin to surpass traditional compression techniques in the aspect of rate-distortion efficiency \cite{balle2016end,cheng2020image}. At the same time, visual applications are also undergoing huge changes, such as image detection \cite{redmon2016you}, recognition \cite{SimonyanZ14a}, and segmentation \cite{chen2017deeplab} tasks. Therefore, it offers possibility to bridge compression and visual analysis together. 
One advantage of using leaning-based methods in image compression is that their compressed domain is \textit{learned} and thus full of compact semantic information, which traditional codecs does not have. Intuitively, compressed data have the purpose to reconstruct the original image so they are supposed to be interpreted as a kind of high-compacted visual representation \cite{torfason2018towards}, which implies the feasibility of directly conducting high-level image processing and machine perception tasks on the compressed domain.

After the pandemic, organizations begin to encourage employees to work from home by using video conferences to interact with coworkers, and when self-quarantining at home short-form video platforms help people kill the time, all of these factors contribute to an explosion of visual data. The question of how to store and transfer this data in an efficient and cost-effective manner has piqued the industry's interest.
Facebook announced their technical solution \cite{Oquab_2021_CVPR} on video chatting by transmitting face landmarks and using generative models under low bandwidth situations. Qualcomm released real-time $1280\times704$ video decoding demo \cite{Qualcomm_realtime} on a smartphone powered by Qualcomm Snapdragon 888 processor. In their work, neural video decoding is supported by AI inference engine and parallelizable entropy coding.
Due to the enormous demand for bandwidth at this time, the relevance of visual integrity at extremely high compression ratios in real-world applications is highlighted.

To bridge the gap between previous research and practical applications, we focus on human face data and present a layered generative compression model that maintains visual fidelity even at extremely high compression ratios. 

The design of the layered end-to-end compression pipeline is fundamentally motivated by Marr's computational theory \cite{Marr1983VisionAC} that vision can be construed as one signal processing system which naturally implicit the support for the layered operation. 
According to the image-to-image series work\cite{huang2018munit,lee2020drit++,choi2018stargan,Isola2016ImagetoImageTW}, image style and texture information can be described by a one-dimensional vector, while content information still needs spatial knowledge whose representation form is two-dimensional. 
Moreover, it is proved that key information can be effectively reserved through layered operation, especially under extremely low bitrate coding scenarios\cite{chang2019layered,chang2021thousand}.

Meanwhile, to directly analyze the compressed data for machine perception tasks, a task-agnostic multi-task analysis model is proposed.

In this paper, our main contributions can be summarized as follows:
\begin{enumerate}
    \item We study and theoretically analyze the characteristics of the machine perception-driven learning-based compression model frameworks. Through these analyses, we present the optimization formulations of those methods and uncover the lack of flexibility of existing layered image compression methods.
    \item We propose a machine perception-driven layered compression model, which is learned in an end-to-end way that each layer is \textit{learnable} and thus semantic-rich. The proposed model serves for both human vision and machine perception tasks that extensive experiments are conducted on CelebAMask-HQ dataset. 
    \item Regarding the coding performance, the proposed model outperforms both traditional codecs (e.g., VVC) and state-of-the-art learning-based codecs in perceptual metrics (i.e., FID\cite{heusel2017gans}, DISTS\cite{ding2020comparison}, and LPIPS\cite{zhang2018unreasonable}), especially at extremely low bit rate condition (i.e., bpp $\le 0.1$). For the semantic analysis tasks, comparable visual analysis performance is achieved on compressed data, saving up to $99.6$\% bit rates in transmission compared with analyzing original RGB images. The multi-task analysis model is proposed to further bring gains on visual tasks, and reduce model complexity.
    
    
    
\end{enumerate}

\begin{figure}[t]
\centering
\includegraphics[width=0.75\linewidth]{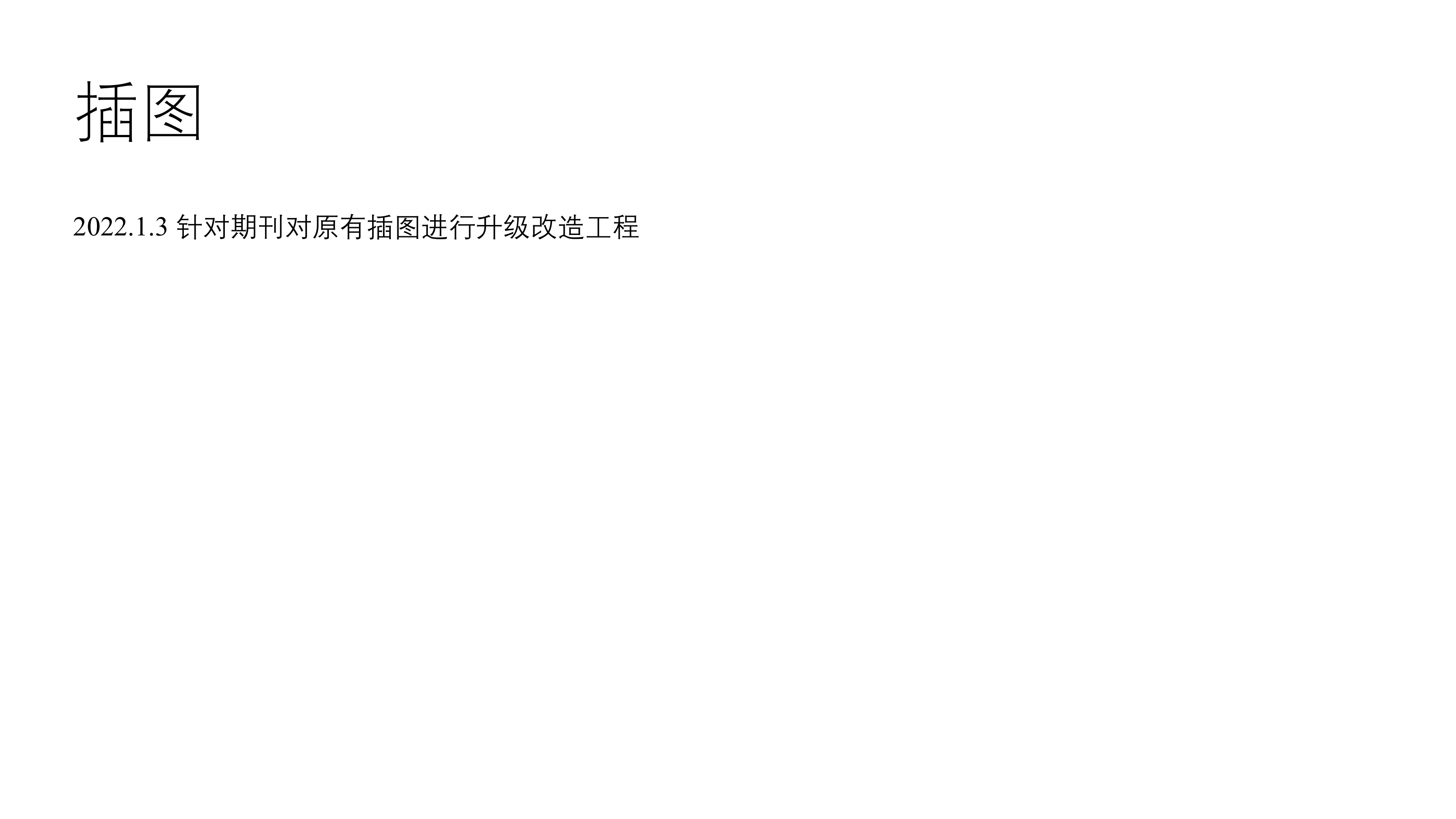} 
\caption{Image compression framework.}
\label{fig:arch1}
\end{figure}

\begin{figure}[t]
\centering
\includegraphics[width=1\linewidth]{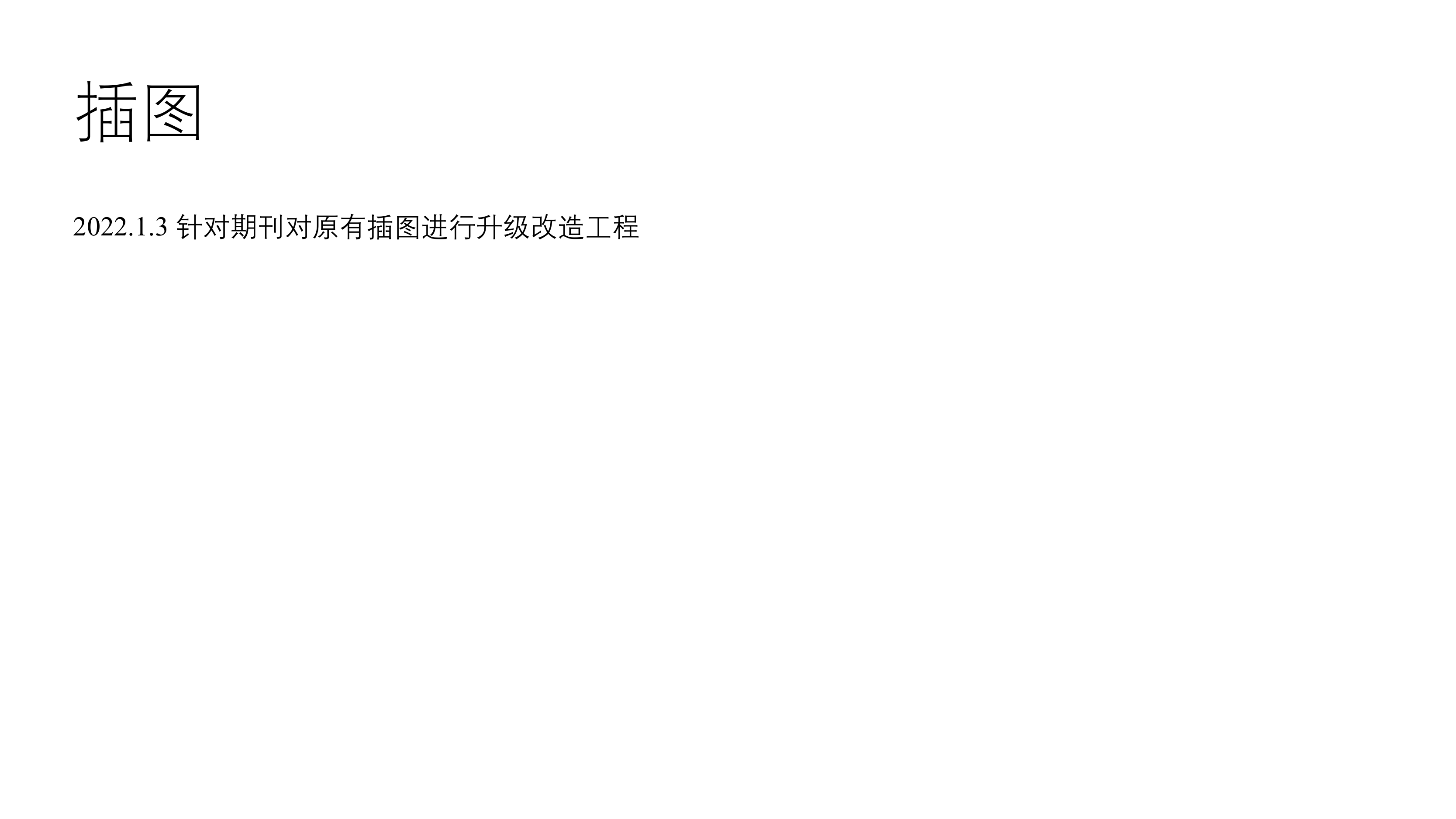} 
\caption{The framework of conducting machine perception on the reconstructed image.}
\label{fig:arch2}
\end{figure}

The remainder of this paper is organized as follows. Section~\ref{sec:Problem} analyzes coding for machine perception and optimization formulations. Section~\ref{sec:Related} reviews the related works of this paper. Section~\ref{sec:Method} introduce the main contribution of the proposed approach by displaying the framework. Section~\ref{sec:Experiments} and Section~\ref{sec:Ablation} show experimental results and discussions based on our model. Finally, conclusions and implications for future work are discussed in Section~\ref{sec:Conclusion}.

\begin{figure}[t]
\centering
\includegraphics[width=1\linewidth]{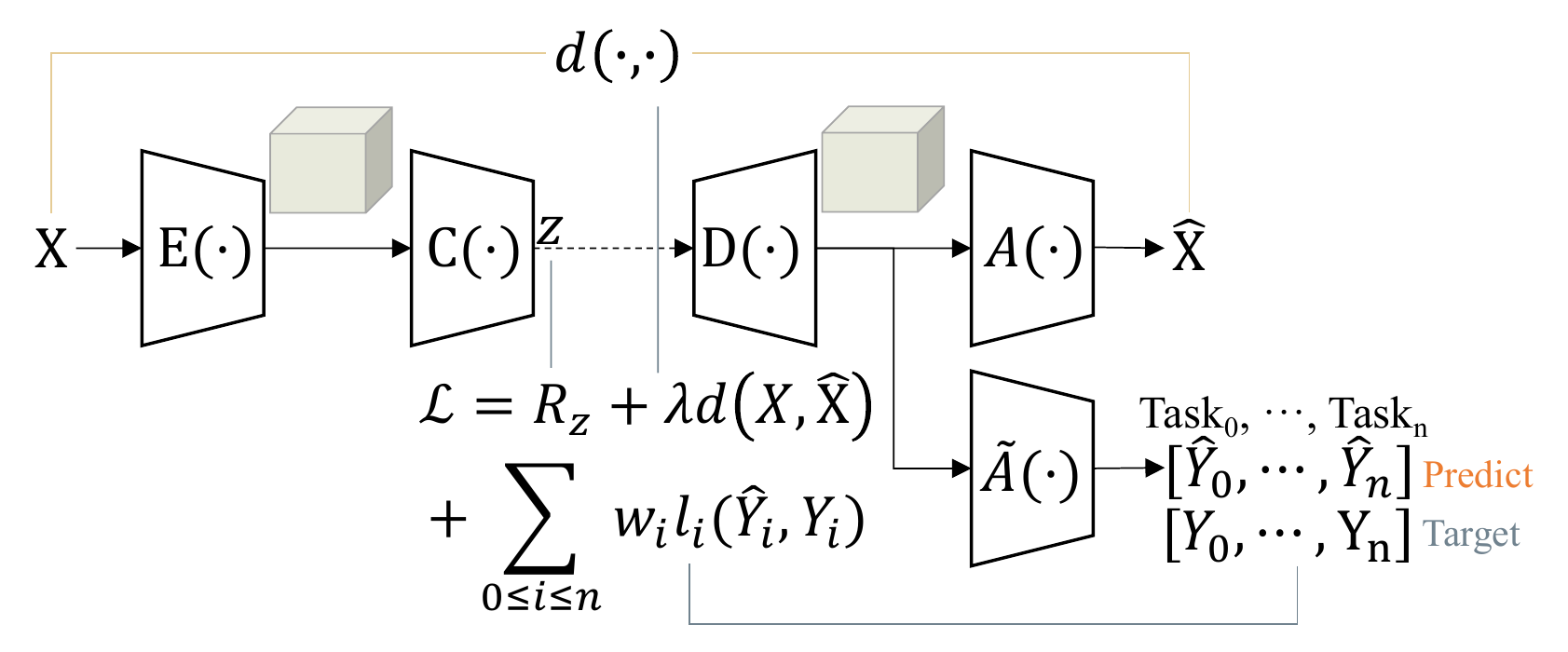} 
\caption{The framework of directly conducting machine perception on the compressed visual data.}
\label{fig:arch4}
\end{figure}

\begin{figure}[t]
\centering
\includegraphics[width=1\linewidth]{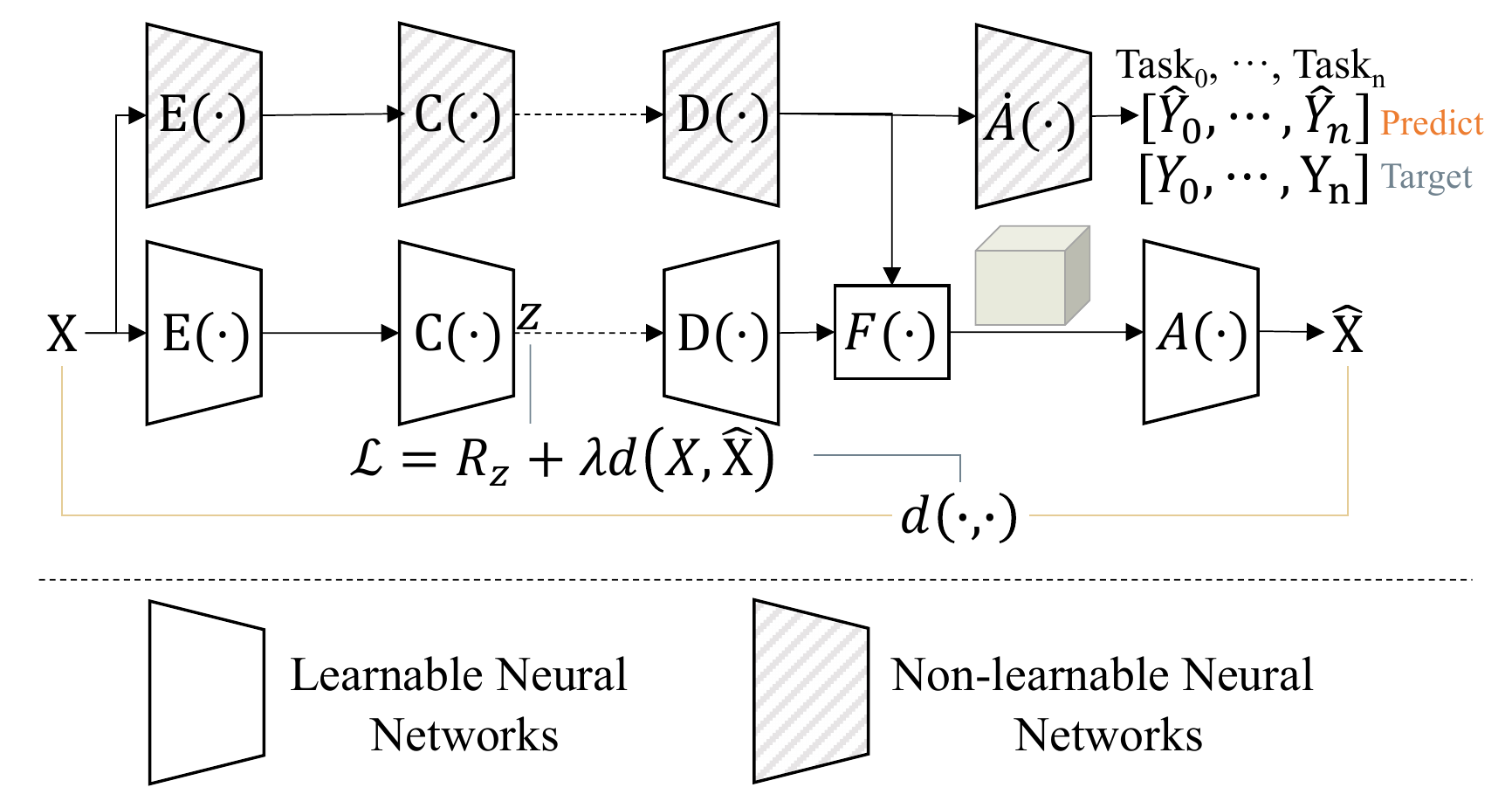} 
\caption{The framework of side information assisted layered image compression.}
\label{fig:arch3}
\end{figure}

\section{Problem Formulation}
\label{sec:Problem}

To illustrate general setups for the machine perception-driven compression models, we define the following components as \cite{Yang2021VideoCF}:
\begin{itemize}
    \item $\mathrm{E}(\cdot | \theta_{\mathrm{E}})$ which transforms visual input to feature matrix;
    \item $\mathrm{C}(\cdot | \theta_{\mathrm{C}})$ which codes features into bit-stream, including probability estimation and entropy coding steps;
    \item $\mathrm{D}(\cdot | \theta_{\mathrm{D}})$ which decodes bit-stream into feature matrix.
\end{itemize}

For different input data types and purposes, we define the following analysis networks:
\begin{itemize}
    \item $\mathrm{A}(\cdot | \theta_{\mathrm{A}})$ that projects features to reconstructed images for human vision purpose;
    \item $\mathrm{A}'(\cdot | \theta_{\mathrm{A}'})$ servers for machine vision tasks, taking reconstruction images as input;
    \item $\Dot{\mathrm{A}}(\cdot | \theta_{\Dot{\mathrm{A}}})$ servers for machine vision tasks, taking hand-crafted side features as input;
    \item $\widetilde{\mathrm{A}}(\cdot | \theta_{\widetilde{\mathrm{A}}})$ servers for machine vision tasks, taking intermediate decoded features as input.
\end{itemize}

We categorize previous related research by their training targets and analyze their corresponding training components in gradient back-propagation. For learning-based image compression methods shown in Fig.~\ref{fig:arch1}, their training target is to balance the trade-off between rate and distortion as:
\begin{equation}
    \mathop{\arg \min}_{\Theta} \mathrm{R}_z + \lambda \mathrm{d},  \quad 
    \Theta = \{\theta_{\mathrm{E}}, \theta_{\mathrm{C}}, \theta_{\mathrm{D}}, \theta_{\mathrm{A}}\}.
\end{equation}

Taking machine perception into consideration, the training target is to find the balance point between rate, distortion, and visual task performance then the optimization function can be formulated as:
\begin{equation}
    \mathop{\arg \min}_{\Theta} \mathrm{R}_z + \lambda d(X, \widehat{\mathrm{X}}) + \sum_{0 \leq i \leq n} w_i l_i(\hat{Y}_i, Y_i),
\end{equation}
\noindent where $X$ is input image, $\hat{X}$ is reconstructed image, $Y_i$ is $i$th task's groundtruth downstream task label, $\hat{Y}_i$ is the prediction results for task $i$, $d$ is image distortion metric and $l_i$ represents the task loss metric for task $i$.
If downstream tasks take reconstructed images $\hat{X}$ as input illustrated as Fig.~\ref{fig:arch2}, visual task prediction results $\hat{Y}$ and optimization parameters are as:
\begin{equation}
\hat{Y} = \mathrm{A}'(\hat{X}), \quad 
  \Theta = \{\theta_{\mathrm{E}}, \theta_{\mathrm{C}}, \theta_{\mathrm{D}}, \theta_{\mathrm{A}}, \theta_{\mathrm{A}'}\}.
\end{equation}

If the analysis model takes intermediate features $Z$ as input shown as Fig.~\ref{fig:arch4}, the optimization function is as:
\begin{equation}
  \hat{Y} = \widetilde{\mathrm{A}}(Z), \quad 
  \Theta = \{\theta_{\mathrm{E}}, \theta_{\mathrm{C}}, \theta_{\mathrm{D}}, \theta_{\mathrm{A}}, \theta_{\widetilde{\mathrm{A}}}\}.
\end{equation}
For separate training, the analysis module is trained by optimizing $\theta_{\mathrm{A}'}$ or $\theta_{\widetilde{\mathrm{A}}}$, independent from compression modules.

\section{Related Work}
\label{sec:Related}

\subsection{Compressed Visual Data for Machine Perception}

There are two types of previous studies on how to analyze compressed representations: traditional codec compressed representations and learning-based codec compressed representations.

From JPEG \cite{wallace1992jpeg} to recent VVC \cite{bross2021overview} standard, traditional codecs can all be considered as a block-based prediction, transform, and quantization. To directly analyze image/video bit-stream from traditional codecs, it is common to take transform parameters \cite{gueguen2018faster,ehrlich2019deep,santos2021less} or intermediate coding blocks \cite{chadha2017video,wu2018compressed} as input. However, in comparison to learned methods, those compressed representations are hand-crafted and lack extensibility.

Learning-based codecs \cite{balle2016end,balle2018variational,agustsson2017soft} directly optimize rate-distortion trade-off commonly using an auto-encoder architecture where quantization, probability estimation, and entropy coding are conducted to achieve high compression ratio. The basic framework for variational auto-encoder (VAE) based compression is illustrated in Fig.~\ref{fig:arch1} where pixel-level distortion loss is used for training. 
With the machine vision attracting more attention, research works \cite{Hou2020Learning,Choi2020TaskAware,shurun2021end} are done to figure out how compression affects machine perception performance when the analysis model's input is a reconstructed image rather than the original one. A high-level view of those methods is shown in Fig.~\ref{fig:arch2} where the reconstructed image $\hat{X}$ is input into vision analysis model $A'(\cdot)$.

In order to efficiently execute machine perception tasks, since the compressed representation of learning-based compression methods can be deemed flexible and semantics-rich, recent researches \cite{torfason2018towards,duan2020video,liu2021semantics,Yang2021VideoCF} have proven the potential of directly conducting visual tasks on the compressed representation without decoding, whose framework is shown as Fig.~\ref{fig:arch4}.
Those methods can be categorized into two groups based on their demand for task-related previous knowledge: task-aware and task-agnostic. 
For the first group, compression models are designed to solve particular downstream tasks \cite{liu2021semantics,alvar2019multi,Yang2021VideoCF,Huang2022HMFVCAH} by training compression and task analysis models together. For the second one, compression works are done without noticing downstream tasks to be solved \cite{torfason2018towards,codevilla2021learned}.

\subsection{Layered Image Compression}
Layered image compression is aided by prior knowledge that the pre-defined information can take several forms, e.g., edge map \cite{chang2019layered}, semantic segmentation \cite{akbari2019dsslic}, face landmarks \cite{Oquab_2021_CVPR}, etc. As shown in Fig.~\ref{fig:arch3}, visual data is first transformed into layered feature components, one of which is hand-designed and the other learned, and then coded separately. Pre-defined layer information is designed to help with specific downstream visual analysis tasks, for example, face edge map can support landmark detection \cite{hu2020towards}.


\begin{figure*}[t]
\begin{center}
\includegraphics[width=0.85\linewidth]{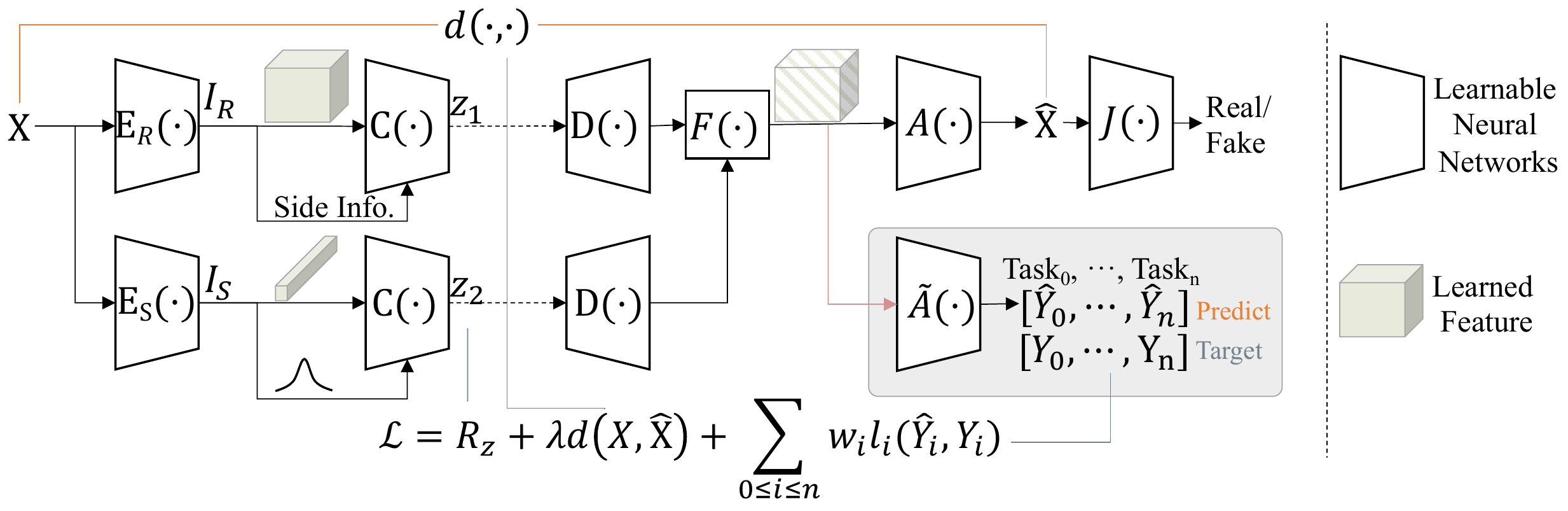}
\end{center}
   \caption{Overview of our proposed approach. We transform the original image signal into two-layered representations: reconstruction-oriented $I_R$ and semantic-oriented $I_S$. Two targets are illustrated, including 1) compression; 2) machine perception of the compressed representation.
   }
\label{fig:model_overview}
\end{figure*}

\begin{figure}[ht]
\begin{center}
  \includegraphics[width=1\linewidth]{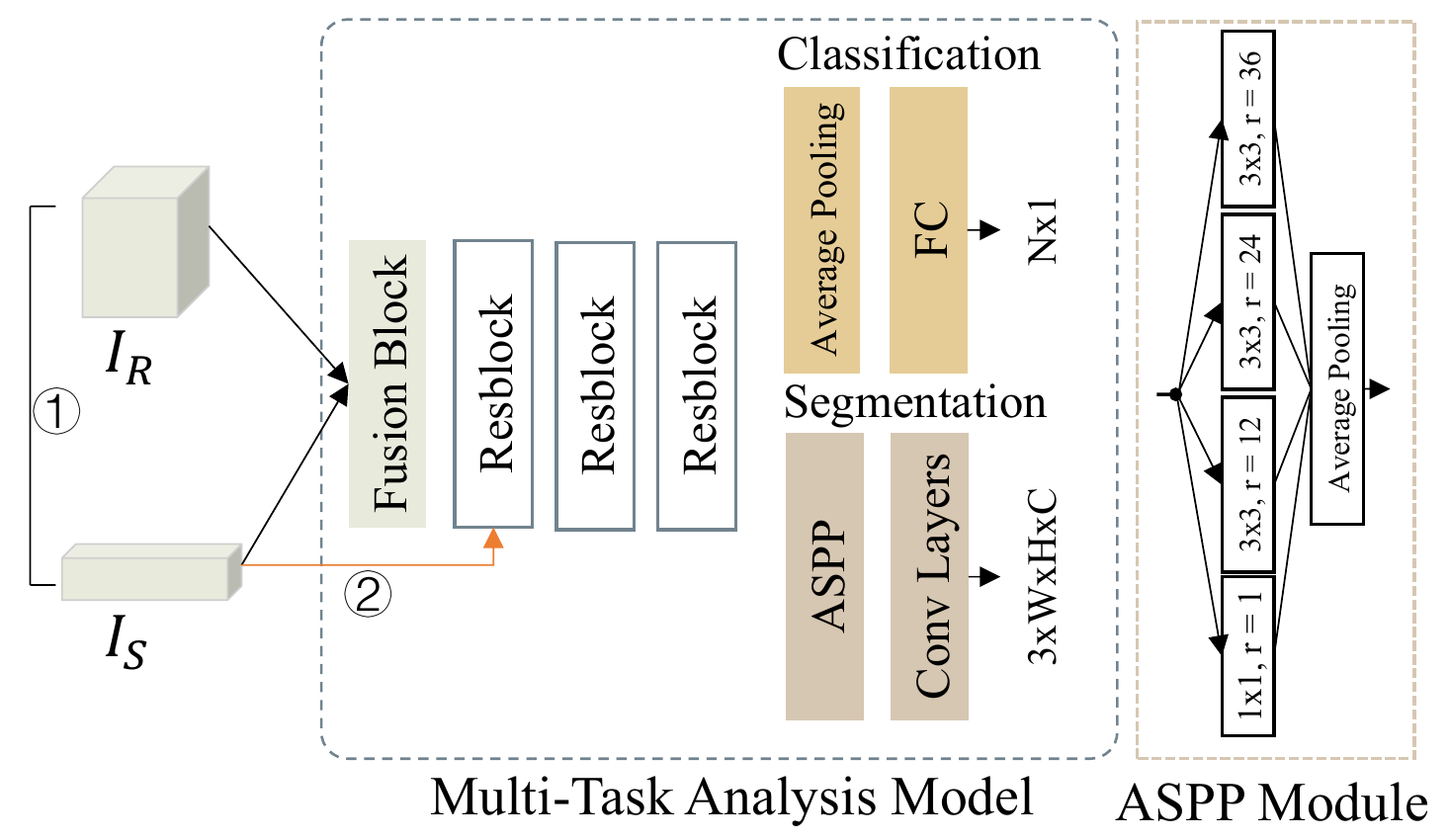}
\end{center}
\caption{Illustration of our multi-task analysis network architecture. Here $C$ and $S$ are class numbers for classification and segmentation, respectively. For the \textit{Resblock}, we follow the second kind of deep residual block design in \cite{he2016deep}.}

\label{fig:multitask_network_structure}
\end{figure}

\begin{figure*}[ht]
\begin{center}
\includegraphics[width=1\linewidth]{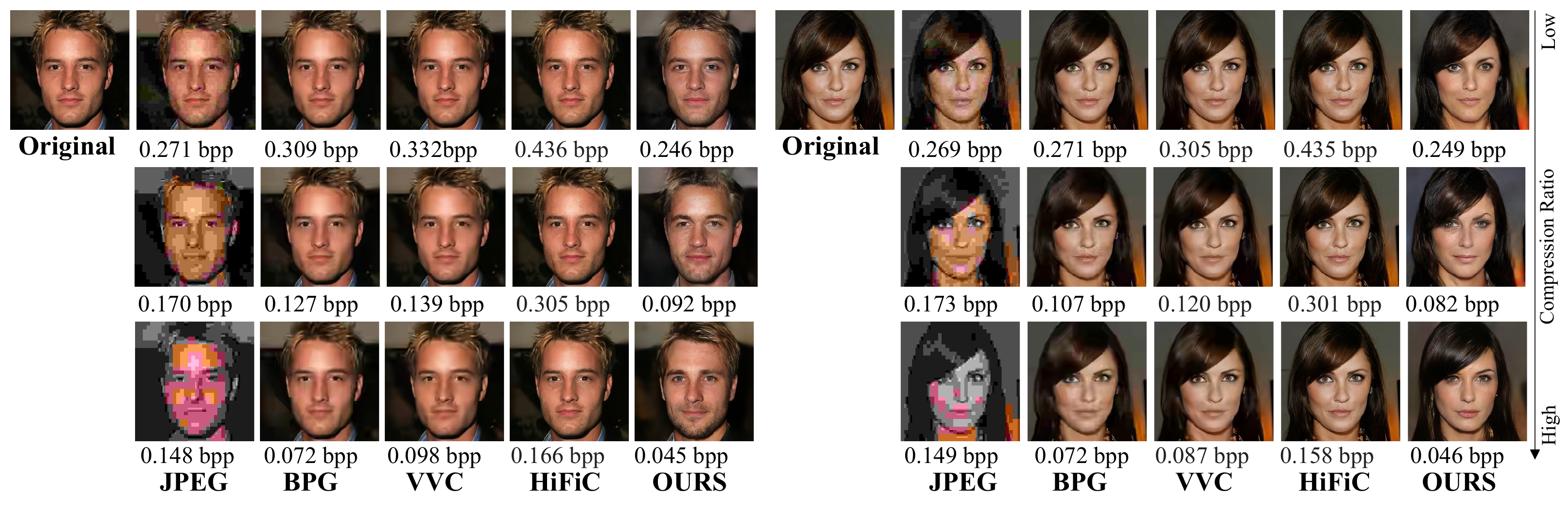}
\end{center}
   \caption{Visual comparison of the reconstructed images with both traditional and learning-based codecs. For each reconstructed image, we display its bit rate (i.e., in bit per pixel, bpp) under itself. Since the resulting bit rates of the compression models are not continuous, we adapt the binary search to find the closest bit rate output for each compression method to ours.}
\label{fig:recon_images}
\end{figure*}

\begin{figure*}[ht]
\begin{center}
   \includegraphics[width=1.0\linewidth]{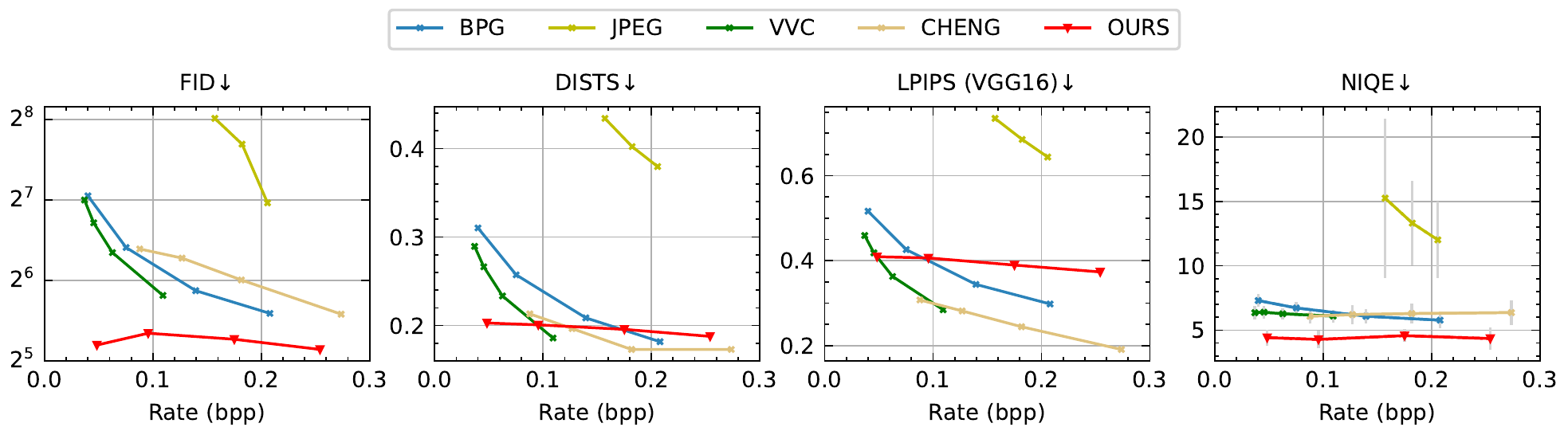}
\end{center}
   \caption{Rate-perception curves of our compression model. Arrow $\downarrow$ in the title indicates the metric value is the lower the better. Grey arrow bars shown in the NIQE metric suggest the data variance while the data variance in the other three metrics is small to be neglected. 
   }
\label{fig:recon_image_metrics}
\end{figure*}

\section{Method}
\label{sec:Method}

The overall flowchart of our proposed model is shown in Fig.~\ref{fig:model_overview}, which involves the compression part and the machine task analysis part.

\subsection{Learning-Based Layered Compression}



Our learning-based layered compression model has three main components: encoder, decoder, and probability estimation model. We first designed a layered encoder and generative decoder to efficiently compress visual data. The optimization goal is then outlined. 

\subsubsection{Layered encoder}

According to the image-to-image series work\cite{huang2018munit,lee2020drit++,choi2018stargan,Isola2016ImagetoImageTW}, image style and texture information can be described by a one-dimensional vector, while content information still needs spatial knowledge whose representation form is two-dimensional. 
Moreover, it is proved that key information can be effectively reserved through layered operation, especially under extremely low bitrate coding scenarios\cite{chang2019layered,chang2021thousand}.

Different encoders are adopted to encode the input image signal $X$ into feature $I_R$ and $I_S$, supporting reconstruction and semantic analysis purposes, respectively. Noting corresponding encoders as $E_R$ and $E_S$ that latent representations can be calculated as:
\begin{gather}
I_R = E_R(X), 
I_S = E_S(X).
\end{gather}

Regarding distinct data distributions, we use independent probability distribution models to further transform $I$ into compact $z$, which is then encoded into bit-streams using entropy encoding methods based on data distribution estimation. 
By calculating cross-entropy $H(p, q)$, which is proven to be the upper bound of $z$'s theoretical bit-stream size \cite{balle2018variational}, the parametric probability model $q$ is adopted to indirectly estimate the probability distribution of $z$ whose probability distribution is $p$. 

Specifically, the hyperprior entropy model \cite{balle2018variational} is utilized for the entropy estimation of $I_R$. 
Latent representation $I_R$ is first quantized to a vector of discrete symbols $
\mathbf{X}=\left\{X_1, X_2, \ldots, X_n\right\}$
. Denoting hyperprior as $\mathbf{Y}$, the conditional probability can be written as $P_{\mathbf{X}|\mathbf{Y}}$. The conditional distribution of each element $X_i$ in $\mathbf{X}$ can be assumed as Gaussian distribution, i.e.,
\begin{equation}
p_{\hat{\boldsymbol{X}} \mid \hat{\boldsymbol{Y}}}(\hat{\boldsymbol{X}} \mid \hat{\boldsymbol{Y}}) \sim \mathcal{N}\left(\boldsymbol{\mu}, \boldsymbol{\sigma}\right)
\end{equation}
The likelihood of the latent representation can be calculated as follows:
\begin{equation}
\begin{aligned}
p_{\hat{\boldsymbol{X}} \mid \hat{\boldsymbol{Y}}}(\hat{\boldsymbol{X}}=\hat{x_i} \mid \hat{\boldsymbol{Y}}) &=  \mathcal{N}\left(\boldsymbol{\mu}, \boldsymbol{\sigma}\right) * \mathcal{U}(-\frac{1}{2},\frac{1}{2})(\hat{x_i}) \\
&= \phi (\hat{x_i} + \frac{1}{2}) - \phi (\hat{y_i} - \frac{1}{2})
\end{aligned}
\end{equation}
where $\phi$ denotes the cumulative function for a standard Gaussian distribution. Non-parametric factorized density model \cite{balle2016end} with Gaussian distribution is adopted for $I_S$ due to its low data dimension setting.

\subsubsection{Fusion module and decoder}
Our decoder works in a generative way with the help of fusion module $F$ to reconstruct the original visual signal. 
Adaptive Instance Normalization (AdaIN) \cite{huang2018munit} is applied as the fusion module that semantic-rich representation $z_2$ can be considered as the input of a multi-layer perception (MLP) predicting the mean and variance for convolution layers in residual blocks, i.e., $F(z_1, z_2) = \operatorname{AdaIN}(z_1, \gamma, \beta)=\gamma\left(\frac{z_1-\mu(z_1)}{\sigma(z_1)}\right)+\beta $, where $\mu, \sigma$ are channel-wise mean and standard deviation, and $\gamma, \beta$ are learnt by MLP taking $z_2$ as the input. An adversarial loss is used for the decoder, producing human-eye-friendly reconstructed images.


\subsubsection{Optimization target}
Training setting is important in bridging data compression and visual analysis. In terms of the task goal, there are two types of training strategies: a) \textbf{for human vision task}; b) \textbf{for machine perception task}. We first introduce the training setup for human vision task here. Then, in the next subsection, we consider the setup for the machine perception task.

Regarding the human vision task, our model's goal is to rebuild the original image signal by predicting pixel values, minimizing the distortion between $X$ and $\hat{X}$ at the same time. In order to satisfy human vision besides preserving signal fidelity, our distortion metric includes three components: the pixel-wise mean-absolute error (MAE) loss $d_{MAE}$, holistic structure-wise SSIM-based loss (i.e., $d_{SSIM} = 1 - SSIM(X, \hat{X})$), and perception-wise loss $d_p$ as \cite{mentzer2020high}. The distortion metric $d$ can be formulated as follows:
\begin{equation}
    d(X, \hat{X}) = \lambda_{MAE} d_{MAE} + \lambda_{SSIM} d_{SSIM} + \lambda_p d_p,
    \label{eqa:distortion}
\end{equation}
where $\lambda_{MAE}$, $\lambda_{SSIM}$, and $\lambda_p$ are hyper-parameters. 
For encoder and decoder parts, training is optimized by:
\begin{gather}
    \mathcal{L}_1 = \mathbb{E}_{X}[-\lambda \log (R(\hat{z})) + d(X, \hat{X})- \beta \log (\mathrm{J}(\hat{X}, z_1))], \label{eqa:L_1}\\
    \Theta = \{\theta_{\mathrm{E}}, \theta_{\mathrm{C}}, \theta_{\mathrm{D}}, \theta_{\mathrm{F}}, \theta_{\mathrm{A}}\}, \label{eqa:L_1_parameters}
\end{gather}

where $\mathrm{J}(\cdot)$ is the discriminator helping generate/decode images meeting the perceptual requirements by telling whether the reconstructed image can be determined as real or fake in the training process. The training loss of $\mathrm{J}(\cdot)$ can be defined as:

\begin{equation}
\begin{aligned}
\mathcal{L}_{\mathrm{J}} = \mathbb{E}_{X}[-\log(1 - \mathrm{J}(\hat{X}, z_1))] + \mathbb{E}_{X}[-\log(\mathrm{J}(X, z_1))],
\end{aligned}
\end{equation}
where $\mathcal{L}_{\mathrm{J}}$ is the loss function for discriminator. 
To be noted, the discriminator $J$ is conditioned on $z_1$ like \cite{mentzer2020high} to obtain sharp image and  we adopt the multi-scale discriminator method proposed by \cite{wang2018pix2pixHD} to eliminate mode collapse problem. Network design details for discriminator $J$ are shown in Table~\ref{table:discriminator_architecture}.

Under this target setting, in the training process, the optimized parameters are $\Theta = \{\theta_{\mathrm{E}}, \theta_{\mathrm{C}}, \theta_{\mathrm{D}}, \theta_{\mathrm{F}}, \theta_{{\mathrm{A}}}\}$ and $\Theta = \{\theta_{\mathrm{J}}\}$, iteratively. In order to use gradient descent methods in the quantization process, uniform noise is added in training, and rounding operation is adopted in the inference stage that values are rounded to their nearest integer, following \cite{balle2016end}. Optimization target setting for machine vision task will be discussed in Section~\ref{sec:visual_analysis}.

\subsection{Machine Perception on Compressed Data}
\label{sec:visual_analysis}
 
Learning-based image compression models have the potential to close the gap between visual data compression and machine perception owing to their flexible and learned latent representation. Visual analysis with compressed data input and the setup for the optimization target will be introduced next.

\subsubsection{Multi-task analysis}
Our hypothesis is that the essential semantics have been embedded and learned in $\hat{z}_1$ and $\hat{z}_2$ that multiple downstream tasks can be solved without decoding. To further improve analysis efficiency, we treat multiple tasks from a multi-task analysis perspective, as,
\begin{equation}
    \hat{Y} = \widetilde{\mathrm{A}}(z_1, z_2), \quad 
    \hat{Y} = [\hat{Y}_0, \cdots, \hat{Y}_n],
\end{equation}
where $n$ is the total number of downstream tasks and $\hat{Y}_i$ is the prediction for $i$th task. 
Classification and segmentation tasks are included as examples because they are indicative of machine perception tasks and have been widely explored, preventing generality loss. The multi-task analysis network details are shown in Fig.~\ref{fig:multitask_network_structure} with task-related modules. One of the most important aspects of multi-task training is learning how to balance distinct task losses\cite{NeurIPS2018_Sener_Koltun}, here we establish hyper-parameters to control the trade-off between tasks, thus the multi-task visual analysis loss can be formulated as,
\begin{equation}
    \mathcal{L}_{\widetilde{\mathrm{A}}} = \lambda_{cls} l_{cls} + \lambda_{seg} l_{seg},
\label{eqa:multitask}
\end{equation}
where $l_{cls}$ and $l_{seg}$ are task-related losses for classification and segmentation, and $\lambda_{cls}, \lambda_{seg}$ are their weight hyper-parameters, respectively. 
In Ablation Studies, we discuss the choice of those hyper-parameters empirically.

\subsubsection{Optimization target}
We investigate two types of training procedures to further explore the relationship between compression and machine perception in order to optimize the compression model for machine perception tasks: separate training and joint training. Different parameters that need to be optimized are taken into account.

In the separate training setup, we fix the compression model and train the multi-task analysis model only, i.e., $\Theta = \{\theta_{\widetilde{\mathrm{A}}}\}$. For joint training, the compression model and the multi-task analysis model are jointly optimized by the total loss and corresponding parameters:
\begin{gather}
    \mathcal{L} = \mathcal{L}_1 + \mathcal{L}_{\mathrm{J}} + \gamma \mathcal{L}_{\widetilde{\mathrm{A}}}, \label{eqa:joint_train_loss} \\
    \Theta = \{\theta_{\mathrm{E}}, \theta_{\mathrm{C}}, \theta_{\mathrm{D}}, \theta_{\mathrm{F}}, \theta_{\mathrm{A}}, \theta_{\widetilde{\mathrm{A}}}\},  \label{eqa:joint_train_parameters}
\end{gather}
where the hyper-parameter $\gamma$ balances the optimized point between human vision and machine perception targets. Specifically, parameters $\theta_{\mathrm{E}}, \theta_{\mathrm{C}}, \theta_{\mathrm{D}}, \theta_{\mathrm{F}}, \theta_{\mathrm{A}}$ are fine-tuned separate train results instead of from scratch following \cite{torfason2018towards} and we set $\gamma$ as $1$ in Section~\ref{sec:Experiments}.

\section{Experiments}
\label{sec:Experiments}

We empirically prove that the main advantage of learning-based image compression methods is not in signal-level preserving but in high perceptual reconstruction and semantic-level information preservation which is feasible for conducting machine perception tasks on compressed visual data.

\subsection{Experimental Settings}

\subsubsection{Dataset}
To evaluate the proposed model, we conduct extensive experiments on CelebAMask-HQ \cite{CelebAMask-HQ} dataset to study the proposed model's performance for human vision and machine perception.
CelebAMask-HQ is a large-scale high-resolution facial semantic dataset that contains $30,000$ face images with resolution $512 \times 512$. Each image is labeled with $40$ attribute classes and has high-quality pixel-level labels of $19$ semantic classes, i.e., $18$ foreground classes and $1$ background classes. We follow the official dataset split setting in \cite{liu2015faceattributes} and report the performance on $256 \times 256$ images.

\subsubsection{Image quality metrics}
Reconstructed images are evaluated by human perception-oriented image quality assessments (i.e., FID \cite{heusel2017gans}, DISTS \cite{ding2020comparison}, LPIPS \cite{zhang2018unreasonable}), which are proposed to rank models for low-level vision tasks according to their perceptual performance. We use the VGG16 c network pretrained on ImageNet \cite{imagenet} as the feature extractor of LPIPS assessment. A no-reference metric NIQE \cite{mittal2012making} is also adopted which measures reconstructed image distribution that deviates from natural image distribution in statics.

\subsubsection{Machine perception metrics}
We verify the feasibility of direct analyzing compressed domain and the effectiveness of multi-task training on two tasks: a) \textbf{multi-attribute estimation} evaluated by calculating average prediction accuracy (Accu.) among $40$ attributes; b) \textbf{semantic segmentation} evaluated by pixel accuracy of all regions, class accuracy for each semantic class, mean Intersection over Union (mIoU) by averaging among all $19$ classes, and Frequency-weighted Intersection over Union (FWIoU) which adds class appearance frequency into consideration on the basis of mIoU.

\subsubsection{Compared Codecs.}
We compare our method's reconstructed image quality with the following baselines of both traditional and learning-based codecs.
Three traditional codecs (i.e., JPEG, BPG, VVC) are taken as the comparison methods. For JPEG, we use cjpeg\footnote{https://linux.die.net/man/1/cjpeg} implementation. BPG is an open-sourced software\footnote{https://bellard.org/bpg/} of H.265/HEVC standard. The VTM\footnote{https://vcgit.hhi.fraunhofer.de/jvet/VVCSoftware\_VTM} version 11.0 is the reference software of H.266/VVC.
For both BPG and VVC, we encode the images using YUV420 color space.
For learning-based compression methods, we have HiFiC~\cite{mentzer2020high} which is based on generative decoding and CHENG \cite{cheng2020image} which achieves comparable performance with VVC regarding subject metric PSNR in comparison.

\begin{table*}[t]

\caption{Classification and segmentation results from the multi-task analysis network. * means that we consider RGB images with $8$ bits per channel which gives $24$ bits per pixel (bpp) in total.
}

\begin{center}
\begin{tabular}{@{}lllllllll@{}}
\toprule
\multirow{2}{*}{Input Type} & \multirow{2}{*}{Codec} & \multirow{2}{*}{\begin{tabular}[c]{@{}l@{}}Rate\\ (bpp)\end{tabular}} & \multicolumn{4}{c}{Segmentation} & \multicolumn{1}{c}{Classi.} & \multicolumn{1}{c}{Quality} \\ \cmidrule(l){4-7} \cmidrule(lr){8-8} \cmidrule(lr){9-9}
 &  &  & mIoU & FwIoU & Accu. & Class Acc. & Accu. & DISTS $\downarrow$ \\ \midrule
Original RGB  & / & 24* & 0.692 & 0.876 & 93.29\% & 78.04\% & 90.59\% & - \\ \midrule
Recon. RGB  & Low & 0.096 & 0.591 & 0.814 & 89.58\% & 69.79\% & 88.12\% & 0.201 \\ \midrule
\multirow{4}{*}{\begin{tabular}[c]{@{}l@{}}Compressed\\ Data\end{tabular}} & Extr. & 0.048 & 0.557 & 0.790 & 88.07\% & 65.90\% & 87.76\% & 0.203 \\
 & Low & 0.096 & 0.572 & 0.805 & 89.01\% & 67.31\% & 88.01\% & 0.201 \\
 & Middle & 0.175 & 0.580 & 0.814 & 89.57\% & 69.73\% & 88.41\% & 0.196 \\
 & High & 0.254 & 0.598 & 0.822 & 90.07\% & 69.73\% & 88.75\% & 0.188 \\ \bottomrule
\end{tabular}
\end{center}

\label{table:multitask_results}
\end{table*}

\subsection{Implementation Details}
\label{section:implementation}

\subsubsection{Separate training schedule}
Separate training schedule are adopted independently for compression and machine perception parts.
For compression part, optimization targets Eq.~(\ref{eqa:L_1_parameters}) and $\Theta = \{\theta_{\widetilde{\mathrm{A}}}\}$ are optimized separately. For the compression part, we use Adam \cite{kingma2014adam} as our optimizer, where $\beta_1$ and $\beta_2$ take the default values of $0.9$ and $0.999$. We adopt a mini-batch size of $8$, and a fixed learning rate of $0.0002$. We follow the Least Squares GAN (LSGAN) training strategy \cite{mao2017least} and use the two time-scale update rule (TTUR) \cite{heusel2017gans} with independent learning rate for the discriminator and decoder/generator to help training convergence. We train the compression model for 3.6M iterations from the scratch. $\lambda_{MAE}$, $\lambda_{SSIM}$, and $\lambda_p$ in Eq.~(\ref{eqa:distortion}) are set as $10$, $0.25$, $0.2$, respectively.

For the machine perception part, we use the same training strategy for single-task and multi-task analysis networks in the separate train. Models are optimized with a stochastic gradient optimizer for $50$ epochs with mini-batch of size $32$. The learning rate starts at $0.001$ and is multiplied by $0.1$ every $10$ epochs. Notably, we do not apply image augmentation methods (e.g., randomly cropping and flipping) because those methods lack semantic meaning in the compressed domain.

\subsubsection{Joint training schedule}
In joint training phase, we jointly optimize compression and multi-task analysis model together by the target loss in Eq.~(\ref{eqa:joint_train_loss}) that the parameters to be optimized are Eq.~(\ref{eqa:joint_train_parameters}). We fine-tune Eq.~(\ref{eqa:joint_train_parameters}) with the pretrained checkpoints from the separate train for $10$ epoches with a fixed learning rate $0.0001$.

\begin{figure}[t]
\begin{center}
   \includegraphics[width=0.9\linewidth]{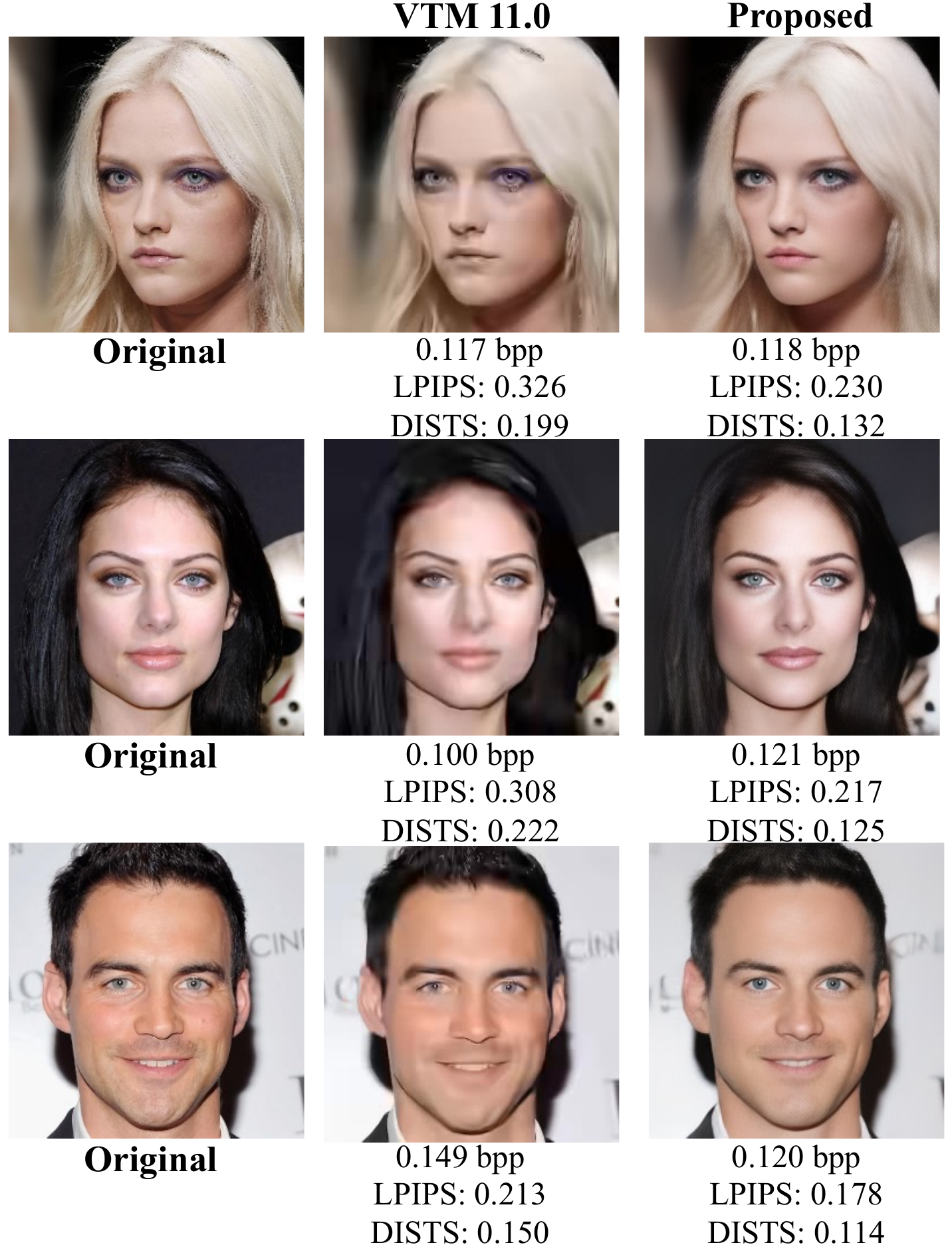}
\end{center}
\caption{
The qualitative comparison results with VVC(VTM 11.0, $\text{qp}=37$) and our method. Lower LPIPS and DISTS indicate better quality.
}
\label{fig:more-images}
\end{figure}

\begin{figure}[t]
\begin{center}
   \includegraphics[width=1.0\linewidth]{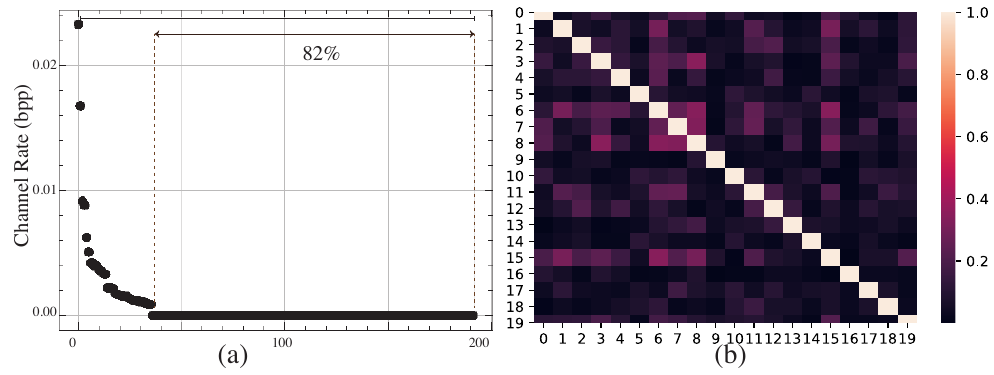}
\end{center}
\caption{Channel-wise visualization analysis results. (a) The rate used for each feature channel is sorted in descending order. (b) Absolute Pearson correlation coefficients among 20 randomly selected channels and consider their absolute value which is in the range $[0, 1]$, the lower value demonstrates the weaker correlation.}
\label{fig:compress_train_strategy}
\end{figure}

\begin{table}[t]
\centering
\caption{
Illustration of the architecture of the discriminator $J(\cdot)$ in the compression model. Convolutional layers are specified with the "Conv" prefix followed by the number of channels, kernel size, normalization, and activation function. 
$\uparrow$ and $\downarrow$ represent upscaling and downscaling, respectively. \textit{IS} is instance normalization and \textit{LN} is layer normalization. \textit{LReLU} represents leaky ReLU activation function.
}
\begin{tabular}{@{}ll@{}}
\toprule
\multicolumn{2}{c}{Discriminator}                               \\ \midrule
Discriminator1                 & Discriminator2                 \\ \midrule
Conv64-4x4, ↓2, LReLU          & Conv32-4x4, ↓2, LReLU          \\
Conv128-4x4, ↓2, IS, LReLU & Conv64-4x4, ↓2, IS, LReLU  \\
Conv256-4x4, ↓2, IS, LReLU & Conv128-4x4, ↓2, IS, LReLU \\
Conv512-4x4, IS, LReLU     & Conv256-4x4, IS, LReLU     \\
Conv1-4x4                      & Conv1-4x4                      \\ \bottomrule
\end{tabular}

\label{table:discriminator_architecture}
\end{table}

\subsubsection{bit rate Control}
Given the fusion module $\mathrm{F}(\cdot)$ used in our design, we propose a straightforward bit rate control strategy to obtain different compression ratio. 
After observing Fig.~\ref{fig:compress_train_strategy}(a), we find out that nearly $80$\% latents’ channels contribute little to the estimated bit rate. Moreover, as depicted in heatmap Fig.~\ref{fig:compress_train_strategy}(b), channel relationships are quite weak, and thus we can represent latent representations by parts of their feature channels.
Based on this analysis, we conduct the bit rate control strategy in which we directly cut down the channel number of the latent representations (i.e., $Z$) to acquire compression models which have the potential to get an extremely high compression ratio. For our compression models from extreme to high bit rate, we set latent representations' channel number in the set $\{8, 16,64, 128\}$, in order.

\subsection{Compression Performance}
From the point of subjective evaluation, we display the reconstructed images in Fig.~\ref{fig:recon_images} to illustrate that our human vision-oriented compression model can maintain vision-friendly information even at extremely low bit rate restriction (i.e., images in the bottom row of Fig.~\ref{fig:recon_images}). 
It is apparent that traditional codecs including JPEG, BPG, and VVC suffer from severe blurring and blocking artifacts. In comparison to VVC baseline, the proposed approach produces noticeably improved reconstruction quality with finer and more accurate texture at two-thirds or less of its bit rate. Although the HiFiC model likewise generates images with great visual quality using image generation methods, it does not investigate the condition of high compression ratio.

From the point of objective evaluation, image quality assessments aiming at reflecting human vision perception including FID, DISTS, LPIPS, and NIQE are illustrated in Fig.~\ref{fig:recon_image_metrics}). Especially on FID and NIQE metrics, our compression model outperforms comparison methods greatly and on DISTS and LPIPS, our model shows advantages at extremely low bit rate circumstances (i.e., bpp $\le 0.1$). 
Objectively, based on the results in Fig.~\ref{fig:recon_image_metrics}), taking DISTS metric as the example and VVC as the anchor,  the proposed method achieves 39.1\%, 27.7\%, 15.1\%, and 6.6\% average reconstruction quality gains under equivalent bit rate (i.e., bpp in the set of \{0.2,0.4,0.6,0.8\}).

\begin{table}[t]
\centering
\caption{Illustration of the gain brought by multi-task training on the classification of compression models with different compression ratio. $\Delta$ is the gain brought by multi-task learning.
}
\begin{tabular}{@{}l|llll@{}}
\toprule
 & \multicolumn{4}{c}{Classification Accuracy (\%)} \\ \midrule
Compression Model & Extreme & Low & Middle & High \\ \midrule
Single-task & 86.74 & 87.74 & 87.76 & 87.82 \\
Multi-task & 87.76 & 88.01 & 88.41 & 88.75 \\ \midrule
$\Delta$ & +1.02 & +0.27 & +0.65 & +0.93 \\ \bottomrule
\end{tabular}

\label{table:multitask_classification}
\end{table}

\subsection{Machine Perception Performance}

\subsubsection{Single-task vs. multi-task}

In Table~\ref{table:multitask_classification}, we explore the potential classification performance gain by multi-task learning analysis network as Fig.~\ref{fig:multitask_network_structure}. Taking classification task as an example, $0.7\%$ accuracy gain is obtained by introducing multi-task learning, compared with single-task learning.

\subsubsection{Semantic analysis}

Semantic analysis results with our multi-task analysis model are displayed in Table \ref{table:multitask_results}.

Here we have four metrics for segmentation task. The results reveal that different metrics reflect different degradation degree and we analyze the cause that our generative learning-based compression model relies seriously on prior knowledge (i.e., data distribution) so it performs worse when a rare accessory appears which harms severely class-related metrics like class accuracy.

\subsubsection{Comparisons with perception on RGB images}
We have the machine perception analysis with RGB image input as the comparison method with our compressed latent input model. Take our low bit rate compression model as an example. In Table \ref{table:multitask_results}, multi-task analysis with its compressed latent as input sacrifices $0.8$\% regarding the mIoU metric of segmentation task meanwhile saves $99.6$\% bit rates (i.e., $(24-0.096)/24 \times 100 \%$) compared with the original RGB images as input. 
After observing experimental results shown in Table~\ref{table:multitask_results}, findings can be concluded as:
\begin{itemize}
    \item Machine perception can be conducted on compressed domain achieving comparable performance with the RGB image input.
    \item Compression ratio is obtained at the cost of reconstructed image quality and machine perception performance.
    \item Visual compressed data contains limited information which shows different analysis effects on different tasks. Pixel-wise segmentation suffers more serious loss than classification task. 
\end{itemize}

\begin{figure}[t]
\begin{center}
   \includegraphics[width=1.0\linewidth]{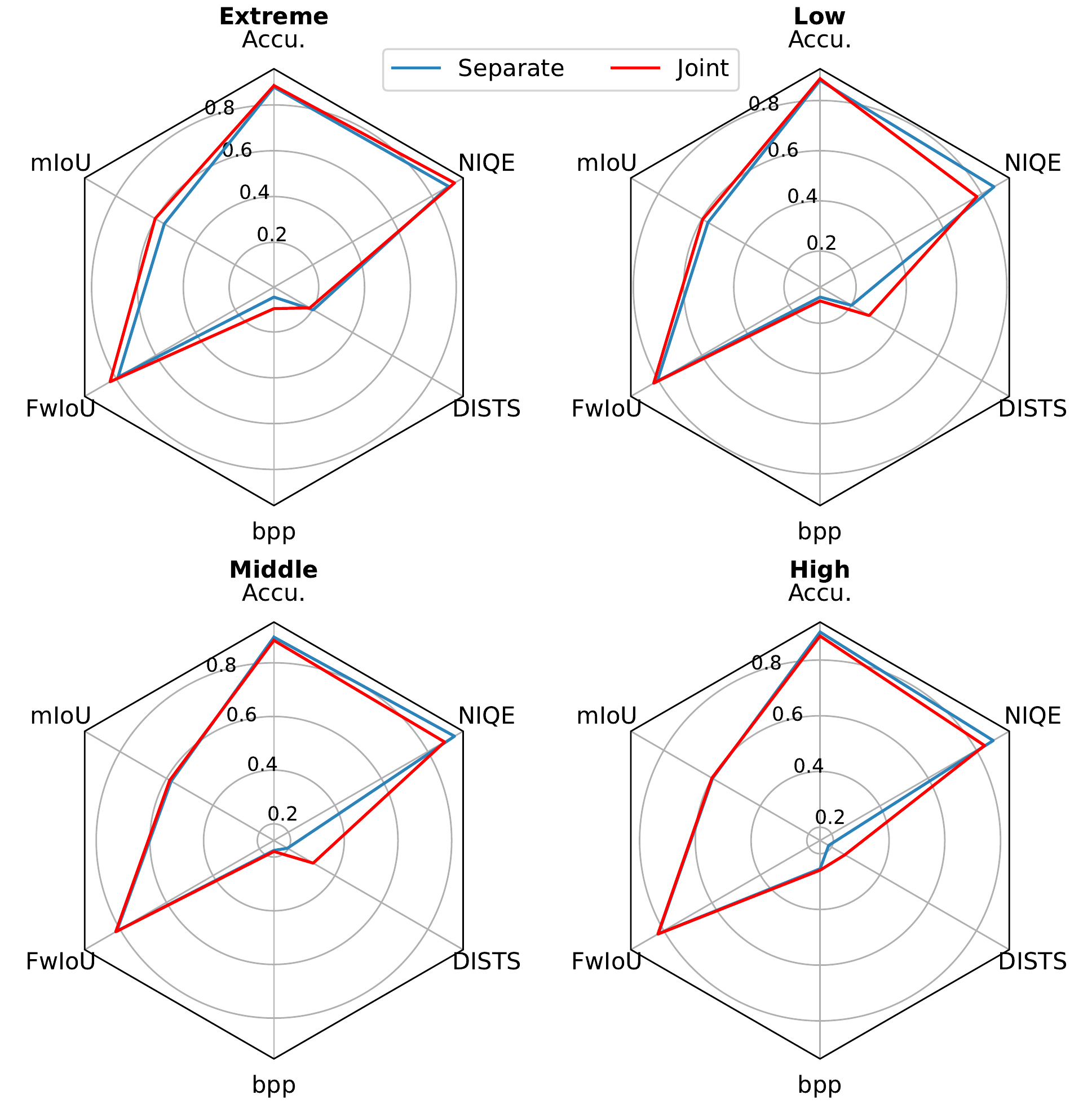}
\end{center}
\caption{The result of joint training schedule compared with separate training on compression ratio, reconstructed image quality, and machine perception performance. The results are from our models of all compression levels. }
\label{fig:radar_compare}
\end{figure}

\subsubsection{Joint training result}

We evaluate the joint training schedule's effect from three dimensions: compression ratio, reconstructed image quality, and machine perception performance. As the radar chart displayed in~\ref{fig:radar_compare}), machine perception performance is elevated at the limited cost of the other two evaluation aspects at extreme compression levels while reconstructed image quality shows trade-off between different quality assessments at high compression level. We display that there exists a machine perception-distortion-human perception trade-off which can be seen as a real-world practical extension for the theoretical work in \cite{liu2019classification}.

\section{Ablation Studies}
\label{sec:Ablation}

\subsection{Model Size}
We declare the feasibility of our proposed model from the prospect of model parameter size. We compare our compressed data input analysis model with representative face attribute prediction models which take the original RGB image as the input. 

As shown in Table~\ref{table:model_complexity_compare}, $50\%$ storage cost on the receiver side can be saved if we directly take compressed data as the input for the downstream tasks. We justify that our model could cost less storage occupation meanwhile save decoding time compared with other RGB image input methods at the decoder side. Direct analysis of the compressed visual data can save $3$\% \#Param compared with analysis on the reconstruction images.

\begin{table}[t]
\centering
\caption{Model parameter size comparison for different face attribute prediction methods. \textit{\#Param} represents the model parameter size. For our compressed data input analysis methods, we also take the encoder model size into consideration, simulating the real-world usage scenario. Our method's results are from the low bit rate compression ratio model.}

\begin{tabular}{@{}lllll@{}}
\toprule
Methods & Encoder & Decoder &  Accu. & \#Param \\ \midrule
Single-task & $\checkmark$ & \xmark &  87.74\% & 26.9M \\
Multi-task & $\checkmark$ & \xmark &  88.01\% & 54.8M \\
Multi-task & $\checkmark$ & $\checkmark$ & 88.12\% & 56.6M \\ \midrule
Liu  \cite{liu2015deep} & \multirow{3}{*}{\xmark} & \multirow{3}{*}{-} & 87.00\% & 100M \\
PSE \cite{zhang2018position} & &  &  91.23\% & 62M \\
DMTL \cite{han2017heterogeneous} &  &  & 92.10\% & 65M \\ \bottomrule
\end{tabular}

\label{table:model_complexity_compare}
\end{table}

\subsection{Codec Hyper-parameters}
Hyper-parameters (usually named $\lambda$) are arbitrarily chosen for achieving different trade-off between distortion and bit rate (bpp) in image compression.
To explore how the hyper-parameters in Eq.~(\ref{eqa:distortion}) effects the rate-distortion trade-off and also machine vision task results, we conduct ablation experiments on $\lambda_{MAE}$ and $\lambda_{SSIM}$. Table~\ref{table:lambda-mae} and Table~\ref{table:lambda-ssim} show the influence of $\lambda_{MAE}$ and $\lambda_{SSIM}$ on the DISTS and classification accuracy, respectively. It can be observed that the best reconstruction and analysis combination is achieved at $\lambda_{MAE}=10$, $\lambda_{SSIM}=1$.

\begin{table}[t]
\caption{
Ablation study for hyperparameter $\lambda_{MAE}$. The best performance is achieved when $\lambda_{MAE} = 10$.
}
\begin{center}
\begin{tabular}{@{}lll@{}}
\toprule
$\lambda_{MAE}$ & DISTS$\downarrow$ & classification (\%) \\ \midrule
100       & 0.230 & 87.01               \\
\textbf{10} & \textbf{0.201} & \textbf{87.76}\\
1         & 0.211 & 87.50               \\ \bottomrule
\end{tabular}
\end{center}
\label{table:lambda-mae}
\end{table}

\begin{table}[t]
\caption{
Ablation study for hyperparameter $\lambda_{SSIM}$. The best performance is achieved when $\lambda_{SSIM} = 1$.
}
\begin{center}
\begin{tabular}{@{}lll@{}}
\toprule
$\lambda_{ssim}$ & DISTS$\downarrow$ & classification (\%) \\ \midrule
1.5         & 0.206    & 87.70               \\
\textbf{1} & \textbf{0.201} & \textbf{87.76} \\
0.75        & 0.203    & 87.71               \\ \bottomrule
\end{tabular}
\end{center}
\label{table:lambda-ssim}
\end{table}

\subsection{Multi-task Analysis Model Loss}
\begin{figure}[t]
\begin{center}
  \includegraphics[width=0.8\linewidth]{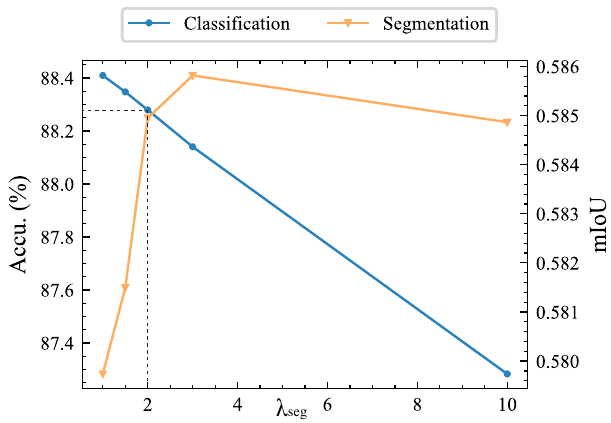}
\end{center}
  \caption{Classification and segmentation tasks' performance by adjusting hyper-parameter $\lambda_{seg}$. The point of intersection has $\lambda_{seg}=2$.}
\label{fig:multitask}
\end{figure}

The relationship among tasks is balanced with hyper-parameters as noted in the loss function Eq.~(\ref{eqa:multitask}). We verify the relationship between those two tasks following the separate training schedule and set weight $\lambda_{cls}$ for the classification as a constant $1.0$, and change task weight $\lambda_{seg}$ for segmentation to see how their performance varies. 
As shown in Fig.~\ref{fig:multitask}, a trade-off relationship between classification and segmentation task is revealed that the best visual analysis performance result is reached when $\lambda_{seg}=2$.

\begin{figure}[t]
\begin{center}
  \includegraphics[width=0.7\linewidth]{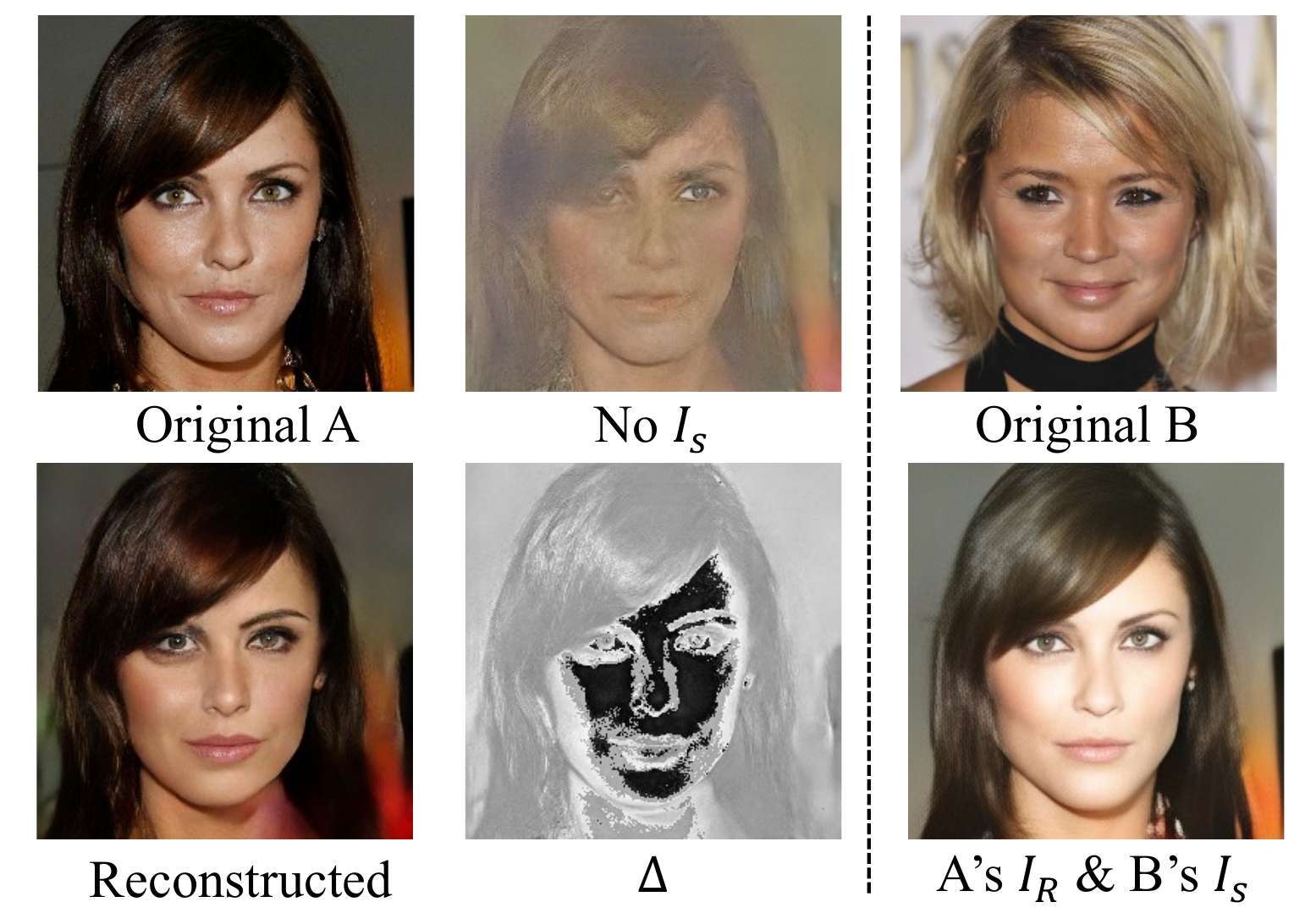}
\end{center}
  \caption{Visualization results if the reconstruction model only has content channel representation as input meanwhile style channel representation is masked with all zeros. $\Delta$ represents for the abstraction results between \textit{Reconstructed} and \textit{No} $I_S$. The bottom right image is the combination of image A's $I_R$ feature and image B's $I_S$ feature.
  }
\label{fig:feature_visualization}
\end{figure}

\begin{figure}[t]
\begin{center}
  \includegraphics[width=1\linewidth]{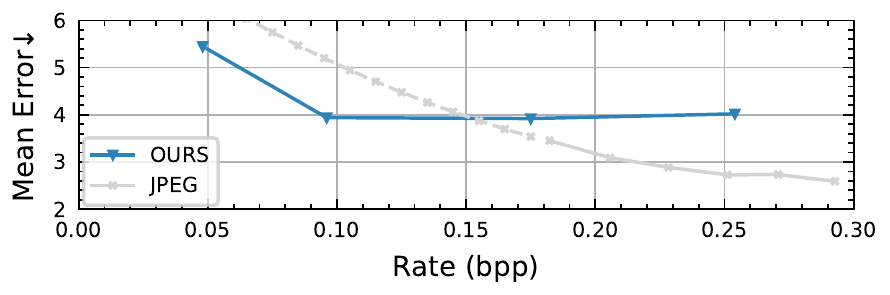}
\end{center}
  \caption{Landmark detection error by the normalized error with inter-ocular distance \cite{cceliktutan2013comparative}. The dashed line is the curve fitting for JPEG under extreme bit rate trending.}
\label{fig:landmark_error}
\end{figure}

\begin{table}[ht]
\caption{1:1 Face recognition results on the reconstructed images of JPEG and the proposed compression method. We set the original image as the anchor and test 1:1 face recognition on the reconstructed image with it. We randomly select 200 images from CelebA-HQ dataset test split. Pass rate is defined as $\sum{T(X, \hat{X}) > \xi}\times 100\%$, where the threshold $\xi$ is set as $0.67$.}

\begin{center}
\begin{tabular}{@{}lll@{}}
\toprule
Codec            & Rate (bpp) & Pass Rate (\%) \\ \midrule
JPEG (qp=3)      & 0.182      & 64.0           \\
JPEG (qp=8)      & 0.293      & 99.5           \\
Proposed (Extr.) & 0.096      & 99.5           \\ \bottomrule
\end{tabular}
\end{center}
\label{table:1-1-face-recognition}
\end{table}

\subsection{Biometric Information Reservation}

To evaluate the accuracy of the reconstructed images with respect to biometric features, we use two image processing tasks: facial feature detection and 1:1 face recognition. The face landmark detection task provides information about the structural change of the face, and the 1:1 face recognition task determines whether two facial images are from the same person or not.

\textbf{Face landmark detection}:
Image compression should produce structure-fidelity and semantic-fidelity reconstructed images for practical use. Considering the characteristic of human face image, we evaluate the structure-fidelity performance by facial landmark detection. We set the detection results on uncompressed images by the pre-trained detection model from \cite{kazemi2014one} as groundtruth landmarks $g$. Here normalized error is defined as:
\begin{equation}
    Error = \frac{\left\|d_{i}-g_{i}\right\|_{2}}{d_{inter}}
\label{eqa:error}
\end{equation}
where $d_i$ is the predicted 2-d locations for landmark $i$ and $d_{inter}$ represents for inter-ocular distance \cite{cceliktutan2013comparative}. \textit{Mean Error} is the mean value of $Error$ in Eq.~(\ref{eqa:error}) for all landmarks.

As illustrated in Fig.~\ref{fig:landmark_error}, our compression model suffers smaller detection error at high compression ratio (i.e., $\le 0.1$ bpp) compared with JPEG, acquiring reconstructed images with more structure information reservation.

\textbf{1:1 face recognition}: We compute the 1:1 similarity rate of face recognition between the original image and the compressed image. A higher similarity rate means less loss of biometric features. We use the online 1:1 face recognition SDK from iFLYTECH\footnote{https://global.iflytek.com/}. The comparison results are shown in Table \ref{table:1-1-face-recognition} that under extreme bit rate constraint, the proposed method can save $67.24\%$ of bit rate to achieve comparable 1:1 face recognition performance on the reconstructed image when taking JPEG codec as the comparison method. It's intuitive that the JPEG undergoes severe blocking artifacts and color shift, which damaged the biometric information of facial images thus reduces the similarity comparison accuracy in 1:1 face recognition. 

\subsection{Hierarchical Representation Effect}

Hierarchical representations can help boost the model's robustness to the quality of visual analysis. We take two approaches to the issue of figuring out how effective hierarchical representations are: 1) \textit{quantitative analysis} in which we conduct ablation tests to determine the impact of hierarchical representations on the outcomes of the visual analysis. 2) \textit{qualitative analysis} to determine the visual effects of latent features following hierarchical decomposition.

\textbf{Quantitative analysis: }We evaluate the effect of hierarchical features by removing the $I_S$ feature, leaving the rest unchanged. We find that at the comparable bit rate, utilizing layered structure (i.e., using both $I_R$ and $I_S$) can result in an average gain of 1.66\% on downstream visual analysis tasks when compared to using non-layered structure (i.e., using $I_R$ only).

\textbf{Qualitative analysis: }
We analyze the effect of each part in the proposed hierarchical representations from the visual viewpoint as shown in updated Figure~\ref{fig:feature_visualization}. We directly show the effect of reconstruction representation $I_R$ by zeroing the semantic 
component $I_S$ as image A. And $\delta$ image is the difference of reconstructed image A and \textit{No} $I_S$, from which we can find out that semantic 
component $I_S$ controls luma and color intuitively from the point of visual effects. 
As defined in Section~\ref{sec:Method}-A, $I_S$ is a one-dimension feature so it does not reflect much information if directly visualizes it. Thus we employ the method that takes $I_R$ from image A and $I_S$ from image B. Then we combine them to see the effect of $I_S$ through the combination image.  From the image in the bottom right corner of Figure~\ref{fig:feature_visualization}, we could find that the luma and color of facial and hair parts are controlled by the target image B.

\section{Conclusion}
\label{sec:Conclusion}

In this paper, we propose an efficient layered generative compression model which achieves comparable downstream machine perception performance with the compressed visual data as input, meanwhile, the reconstructed image is human visual fidelity proved with four perceptual quality metrics. A task-agnostic perception model with compressed data input is developed to achieve analysis efficiency and flexibility, and it efficiently supports diverse analytical tasks. Thorough experiments on face images are conducted verifying that multiple independent visual analysis tasks can be directly reasoned from the compressed representations, meanwhile saving the decoding process and decoder-side computation complexity which demonstrates the proposed model's practical value. 

The analysis and observation in this paper provide feasible directions for conducting visual tasks on latent representations, particularly which are learned. Although learning-based compression models have their limitations for pixel-fidelity-based metrics (e.g., PSNR and MSE), we argue that the principal advantages of learning-based image compression models lay in perception retention and flexible analysis support for machine perception, instead of signal-level fidelity preservation, which shows a direction for future codecs.

\section{References}
\bibliographystyle{IEEEtran}
\bibliography{refs}

\end{document}